# Visualization of nonlinear optics in a microresonator

**Authors**
Hao Zhang[1, 2†], Haochen Yan[1, 3†], Alekhya Ghosh[1, 3], Shuangyou Zhang[1,4], Toby Bi[1, 3], Yaojing Zhang[5], Lewis Hill[1], Jolly Xavier[1, 6], Arghadeep Pal[1, 3], Yongyong Zhuang[1, 7], Jijun He[2], Shilong Pan[2, 8], Pascal Del'Haye[1, 3, 9]

**Affiliations**
[1]*Max Planck Institute for the Science of Light, Staudtstr. 2, 91058, Erlangen, Germany*
[2]*National Key Laboratory of Microwave Photonics, Nanjing University of Aeronautics and Astronautics, Nanjing 210016, China*
[3]*Department of Physics, Friedrich Alexander University Erlangen-Nuremberg, 91058, Germany*
[4]*Department of Electrical and Photonics Engineering, Technical University of Denmark, Kgs. Lyngby 2800, Denmark*
[5]*School of Science and Engineering, The Chinese University of Hong Kong (Shenzhen), Guangdong 518172, China*
[6]*SeNSE, Indian Institute of Technology Delhi, Hauz Khas, 110016, New Delhi, India*
[7]*Electronic Materials Research Laboratory, Key Laboratory of the Ministry of Education & International Center for Dielectric Research, School of Electronic Science and Engineering, Xi'an Jiaotong University, Xi'an, 710049, China*
[8]pans@nuaa.edu.cn
[9]pascal.delhaye@mpl.mpg.de
[†]These authors contributed equally to this work

## Abstract
A precise understanding of nonlinear optical phenomena in whispering gallery mode (WGM) microresonators is crucial for developing next-generation integrated photonic devices. Applications include on-chip sensors for biomedical use, optical memories for all-optical networks and frequency combs for optical clocks. However, our ability to spatially localize nonlinear optical processes within microresonators has been limited because optical feedback is often only collected through a bus waveguide. In this study, we present the direct visualization of nonlinear optical processes using scattering patterns captured by a short-wave infrared (SWIR) camera. Through systematic analysis of these scattering patterns, we can distinguish between different nonlinear effects occurring within the microresonator. Direct imaging of nonlinear processes in microresonators can significantly impact many applications, including the optimization of soliton frequency combs, real-time debugging of photonic circuits, microresonator-based memories, and chip-based data switching in telecom circuits.

## Main
Whispering gallery mode (WGM) microresonators [1-4] have become one of the most extensively studied platforms in recent decades due to their unique nonlinear properties, which enable the observation of numerous phenomena such as frequency comb generation [5-9], parity-time symmetry breaking [10-12], Brillouin scattering [13-19], and many more [20-28]. Leveraging the compact size of WGM microresonators, researchers have achieved a wide range of applications [29-32], including ultra-sensitive biomedical sensing [33-35], compact integrated lasing [36-38], and high-capacity optical communication [30, 39, 40].

However, investigations of cavity phenomena typically rely on indirect measurements using transmission signals from coupling waveguides, which average over spatial information and thus fail to localize intra-cavity nonlinear dynamics [41-43]. Recently, real-time imaging [44-46] has emerged as a promising method to overcome these limitations by enabling direct observation of cavity mode scattering distributions. Real-time visualization of different nonlinear processes inside of microresonators has not been reported in previous work.

While specific applications such as monitoring optical fiber fabrication have been proposed and reported [47], there is no study on direct imaging of nonlinear effects in microresonators. In this work, we present a detailed and systematic study of visualizing various nonlinear optical processes via imaging of scattering patterns in optical ring resonators. Specifically, we uncover unique spatial mode "fingerprints" corresponding to linear and different nonlinear processes including four-wave-mixing (FWM) generation, Q-factor changes induced by a Tungsten tip [48], and stimulated Brillouin scattering (SBS) generation. We quantify these fingerprints and track their evolution under different laser detunings. This data enables us to distinguish different optical processes from image analysis. Our work not only provides additional spatial information for understanding intra-cavity dynamics but also creates a new perspective for the future characterization of more complex integrated photonic circuits.

# Results
## Experimental setup and procedure

We utilize a short-wave infrared (SWIR) camera to capture scattered light from a fused silica microresonator under different conditions such as four-wave mixing (frequency comb generation) and stimulated Brillouin scattering (SBS). A similar setup has been used previously to analyze standing wave patterns in microresonators without optical nonlinearity [49]. Here, we study three different nonlinear optical effects and their impact on scattering patterns, as shown in Fig. 1. Additional details on the experimental setup can be found in the supplementary information (SI).

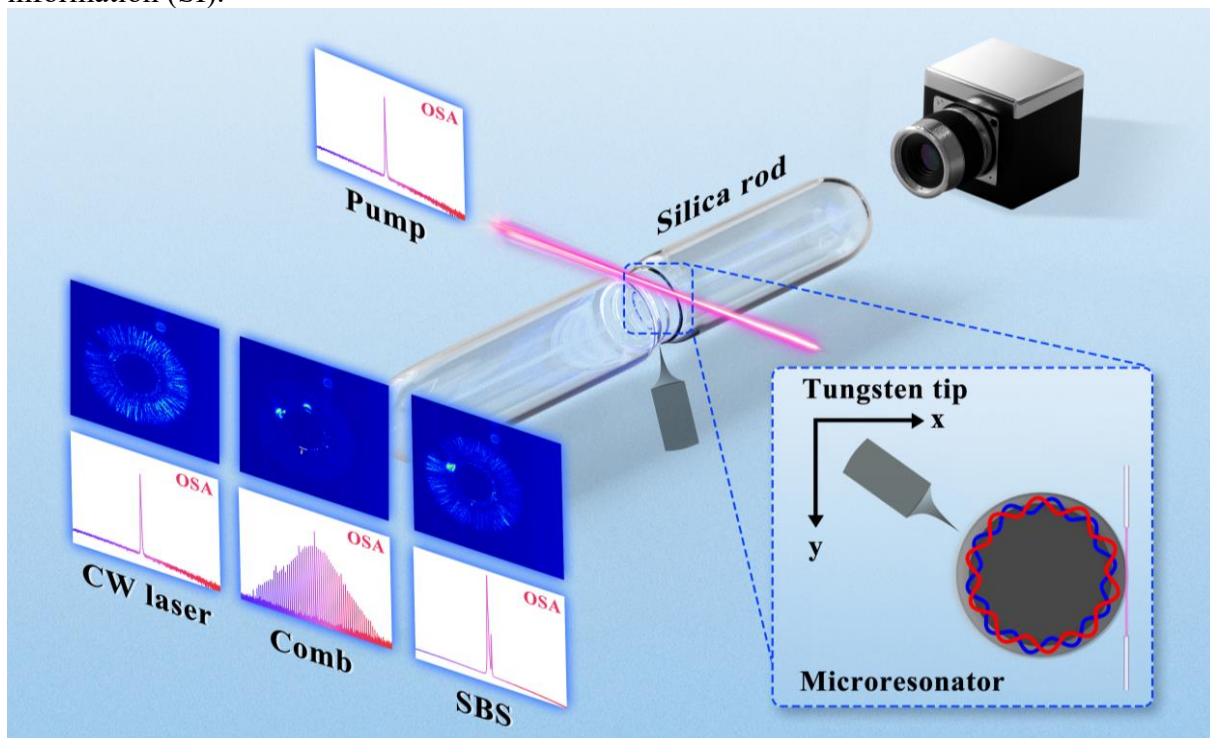

**Figure 1. Schematic of visualizing various nonlinear processes inside a microrod resonator.** Depending on the intracavity nonlinear processes, different scattering patterns are recorded. These scattering patterns are compared to conventionally measured data that is collected via an optical waveguide coupled to the microresonator. The orange arrows denote the position of the tungsten tip, whereas the red arrows indicate the location of the tapered fiber.

## Visualization and characterization of comb states

We first investigate the scattered light associated with dissipative soliton frequency comb states. Specifically, we select three states of interest: a no-comb state, a stable comb state, and a comb-suppressed state, which are achieved by introducing a near-field interaction with a Tungsten tip. These three states are acquired under the same coupling conditions to the microresonator. The no-comb state and stable comb state are achieved by pumping two different resonances, while the "comb-suppressed state" is achieved by pumping the same resonance of the stable comb state. Initially, we record the transmission spectra for both the comb and no-comb states, as illustrated in Fig. 2(a) and Fig. 2(b), where blue denotes transmission and green denotes reflection, a convention maintained throughout this paper. The comb state is generated by pumping Resonance B at 1564.27 nm, while Resonance A at 1564.28 nm cannot generate frequency combs. In the following, we tune an external laser into the two resonances while recording scattering patterns at various detuning values. The resonance depicted in Fig. 2(a), exhibits no nonlinear optical effects. In contrast, the resonance shown in Fig. 2(b), shows both four-wave mixing (FWM) and comb generation. For both resonances, we capture the scattered light with a SWIR camera (see SI: Experimental setup). Selected images at different detuning points of no-comb/comb states are presented in Fig. 2(d) and Fig. 2(e). For comparison, images are captured at identical transmission power levels, labelled as *i, ii, iii* across all examined resonator states. Additionally, we observe a ring-shaped distribution of light, which resembles the circumference of the microresonator. We can see distinct differences between the comb/non-comb case in the captured images. In the presence of a frequency comb, we observe localized scattering peaks in Fig. 2(e), while for the state without comb, we observe noisy patterns as shown in Fig. 2(d). Moreover, a larger amount of light is scattered out of the microresonator in the state without comb, which can be seen from the "halo" around the resonator shown in Fig. 2(d). We present two spectra in Fig. 2(g) and Fig. 2(h), corresponding to the scattering patterns shown in Fig 2(e) *ii* and *iii*. Fig. 2(g) shows a spectrum with the onset of parametric oscillations with multiple optical sidebands when tuning the laser into resonance. Further into the resonance, we reach a chaotic modulational instability comb as shown in Fig. 2(h). Overall, we observe that the comb-generating mode family exhibits fewer scattering losses.

In order to control the amount of nonlinear light conversion in the resonator, we use a sub-wavelength-sized Tungsten tip that changes backscattering in the WGM resonator and reduces the Q-factor [48] (see SI: Scattering pattern induced by Tungsten tip). By adjusting the position of the tungsten tip, we can control both the amplitude and phase of the induced backscattering. Fig. 2(c) shows the changed resonator transmission spectrum in the presence of the tungsten tip. We observe that the fast oscillations in the transmission spectrum of the FWM state are suppressed. Corresponding camera images are shown in Fig. 2(f). The optical spectra from corresponding to Fig. 2(a) and Fig. 2(c) are similar. However, the scattered light patterns in Fig. 2(f) still resemble those in Fig. 2(e) rather than those in Fig. 2(d), which suggests that the FWM suppression does not change the spatial profile of the cavity mode. We can still identify localized scattering at the same locations. In the presence of a frequency comb, the localized scattering regions are shown in Fig. 2(e). The loaded Q-factor of the comb-generating resonance is $1.13\times10^8$. When perturbing the near-field with the tungsten tip, the loaded Q-factor decreases to $0.63×10^8$, which is close to the loaded Q-factor of the "no-comb" resonance ($Q = 0.68×10^8$). The comparison between different scattering patterns enables us to estimate the circulating field intensity without measuring transmission spectra.

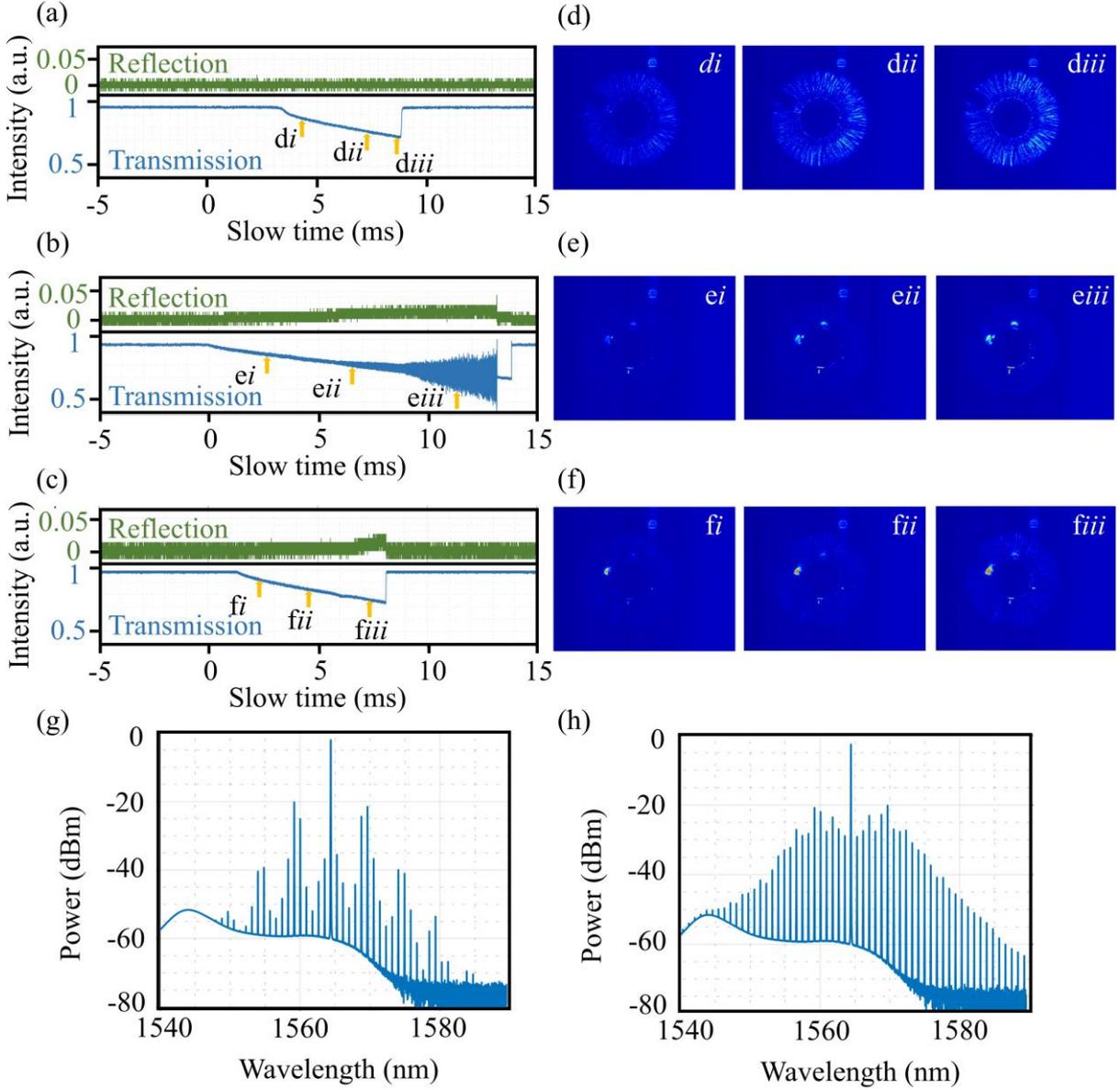

**Figure 2. Identification and visualization of scattered light for different resonator states.** **(a)-(c)** Transmission (blue) and reflection (green) spectra for **(a)** state without nonlinear effects, **(b)** frequency comb state, and **(c)** comb generation suppressed with tungsten tip. We can observe a noisy regime of FWM in Fig. 2(b), while for the exactly same coupling position, a Tungsten tip suppressed the FWM noise as shown in **Fig. 2(c)**. **(d)–(f)** Selected SWIR images at different detunings corresponding to **(a)-(c)**. The sub-index *i-iii* indicates different detuning values. Images recorded at the same transmitted powers are marked with the same sub-index. We observe a quite homogenous distribution of the scattered light for the state without nonlinearity as shown in **Fig. 2(d)**. The images of the nonlinearity-suppressed state in **Fig. 2(f)** are similar to the images of the comb state in **Fig. 2(e)**, although the optical spectrum does not show comb sidebands. **(g)-(h)** Optical spectra corresponding to captured images of panel **2(e)***ii* and panel **2(e)***iii.*

In the following, we try to find a relationship between scattering patterns and different nonlinear processes within the microresonator. To quantitatively analyze the processes, we plot the scattered light intensity for all three states as a function of transmitted optical power, as shown in Figure 3. When we tune into the resonance, more power is coupled to the microring resonator, and less power is transmitted through the tapered fiber coupler. Fig. 3(a) shows the scattering pattern of a "no-comb" state. In this state, the continuous-wave light within the microresonator generates a homogeneous scattering pattern without strong localized scatterers. To estimate the total scattered power, we analyze the scattered light intensity all over the ring region, with the result being shown in Fig. 3(b). We can see a clear linear relation between the scattering intensity and transmitted power (and thus intracavity power). Note that we measure transmitted power, which decreases when the intracavity power increases. In the presence of a frequency comb, we analyze the intracavity power through the localized scatterers shown in Fig. 3(c). We have selected several of these scatterers as regions of interest, each marked by rectangular boxes in different colors. These regions of interest are kept the same for the measurements with the tungsten tip. Unlike the "no-comb" state, the comb state shows a distinct nonlinear trend, as shown in Fig. 3(d)-(f). The nonlinear change in the transmitted power is a result of the onset of nonlinear frequency conversion, which distributes light into other cavity modes. Moreover, the intensity changes across different scatterers. The measurement with the SWIR camera depends on the overlap between a scatterer's physical position with the backscattering-induced standing wave in the microresonator. The nonlinear change in scattering power is attributed to different optical losses in the resonator modes of the comb sidebands. In addition, backscattering of light will generate standing wave components in the intracavity fields. The maxima of these standing waves will shift when changing the detuning through input power changes. This leads to more complex behaviors of the scattering power, depending on the position of the standing wave maxima.

The nonlinear change of the scattered light intensity with coupled power can help to show the presence of nonlinear processes within the microresonator. For further analysis, we measure the total scattered light observed around the ring resonator. Specifically, we fit the ring region for the three states as shown in Fig. 3(g) (no-comb), Fig. 3(h) (comb), and Fig. 3(i) (comb-with-tip) at a specific detuning. We define an "inner" region (red ring) as the region of interest and an "outer" region (yellow ring) that includes additional scatterers. The camera images of these regions are shown in Fig. 3(g) to Fig. 3(i). Subsequently, we plot the scattering distribution along the resonator as a function of detuning, shown in Fig 3(j) to 3(l). The labeled arrows indicate the corresponding positions in detuning for these selected images in Figure 2. The "no-comb" state (Fig. 3(j)) exhibits a different scattering pattern compared to the FWM and FWM-suppressed states (Fig. 3(k) and 3(l)). In the "no-comb" state, the scattered light intensity appears relatively uniform. On the other hand, the scattered light distribution for the comb state shows more localized scatterers.

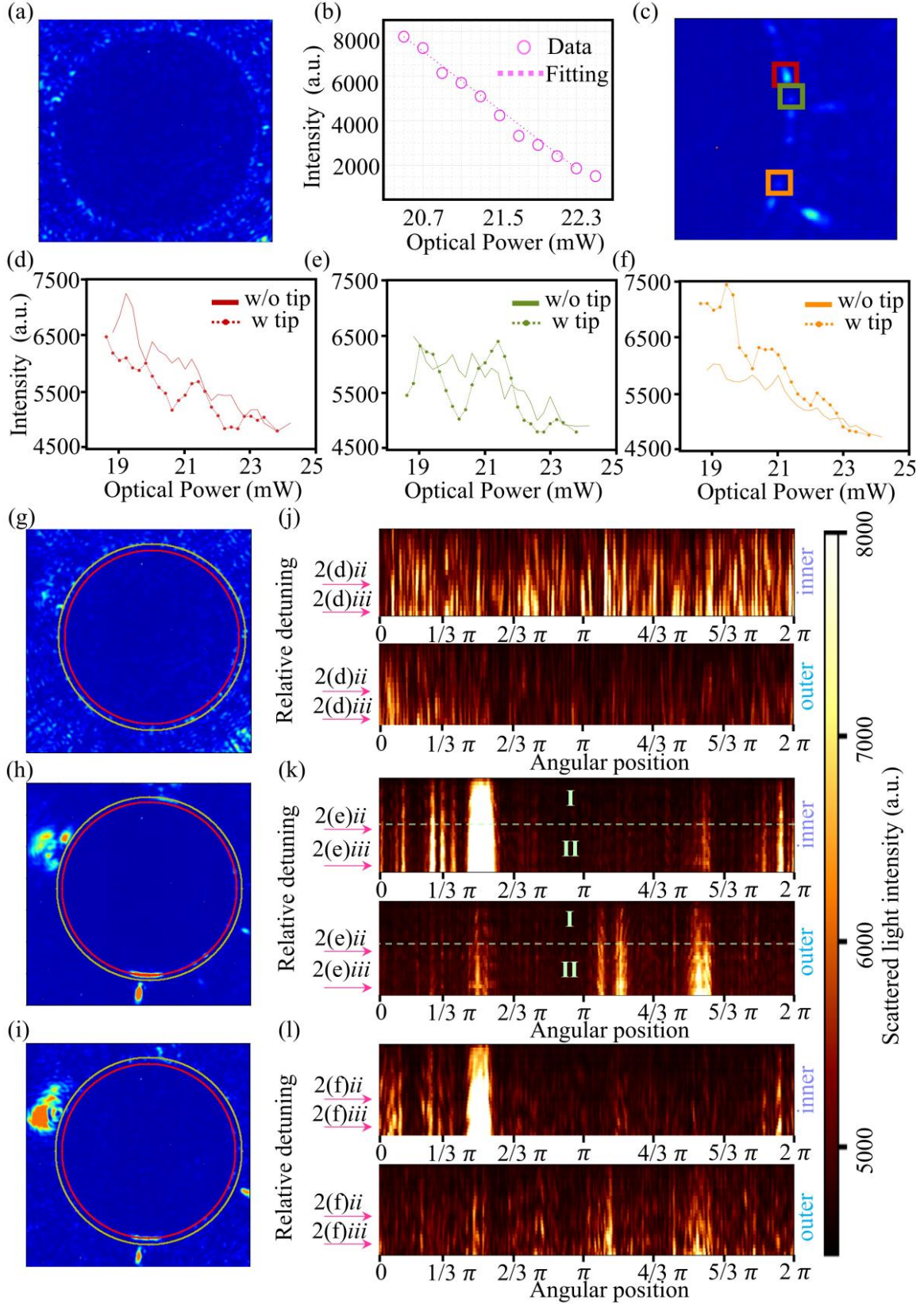

**Figure 3. Analysis of scattered light for states with and without frequency comb. (a)** SWIR image for the state without nonlinearity. **(b)** Total scattered light intensity from the resonator as a function of optical power. **(c)** SWIR image for the comb state. Different localized scattering regions are highlighted with rectangular boxes. The scattering intensity in these boxes is shown in panels **(d) - (f)**. Scattering images of the microresonator for three different states: **(g)** state

without nonlinearity, **(h)** comb state, and **(i)** comb generation suppressed by Tungsten tip. Scattering intensity color maps as functions of angular position and detuning corresponding to **(g)** state without nonlinearity, **(k)** comb state and **(l)** nonlinearity-suppressed state. The outline of the resonator inner ring can be seen as the red circle, while the outer resonator ring can be seen as the yellow circle. The pink arrows shown on the left side of the color maps correspond to the images captured at specific detunings shown in **Fig. 2**.

## Characterization of cascaded Brillouin Scattering

In this section, we investigate nonlinear processes within the microresonator when two different nonlinear processes occur simultaneously. Specifically, we choose a state, in which stimulated Brillouin scattering (SBS) and frequency comb generation happen at the same time. The generation of a similar SBS comb has been investigated in previous work [50]. Fig. 4(a) illustrates the transmission spectrum as a function of detuning, highlighting three resonances, each capable of generating SBS. We label these resonances as Resonance 1, Resonance 1' and Resonance 2. Resonance 1 and Resonance 1' are the same cavity mode, however with different coupling position of the tapered fiber. By tuning the laser into Resonance 1, we can observe first and second order SBS. By tuning the laser frequency into Resonance 1', we can only observe first-order SBS. Resonance 2 belongs to another mode family. When we tune the laser frequency into Resonance 2, we initially observe a first-order SBS sideband and at smaller detunings FWM in addition to the SBS. We first measure the optical spectra of the above-mentioned nonlinear processes as shown in Fig. 4(b), for pump-only, SBS, and SBS + comb. We use the SWIR camera to visualize the scattered light from the microresonator in the presence of SBS, where Fig. 4(c) corresponds to Resonance 1, Fig. 4(d) corresponds to Resonance 1', and Fig. 4(e) corresponds to Resonance 2. Similar to our data in Figure 2, the panels with identical Roman numbers are measured at roughly the same intra-cavity power.

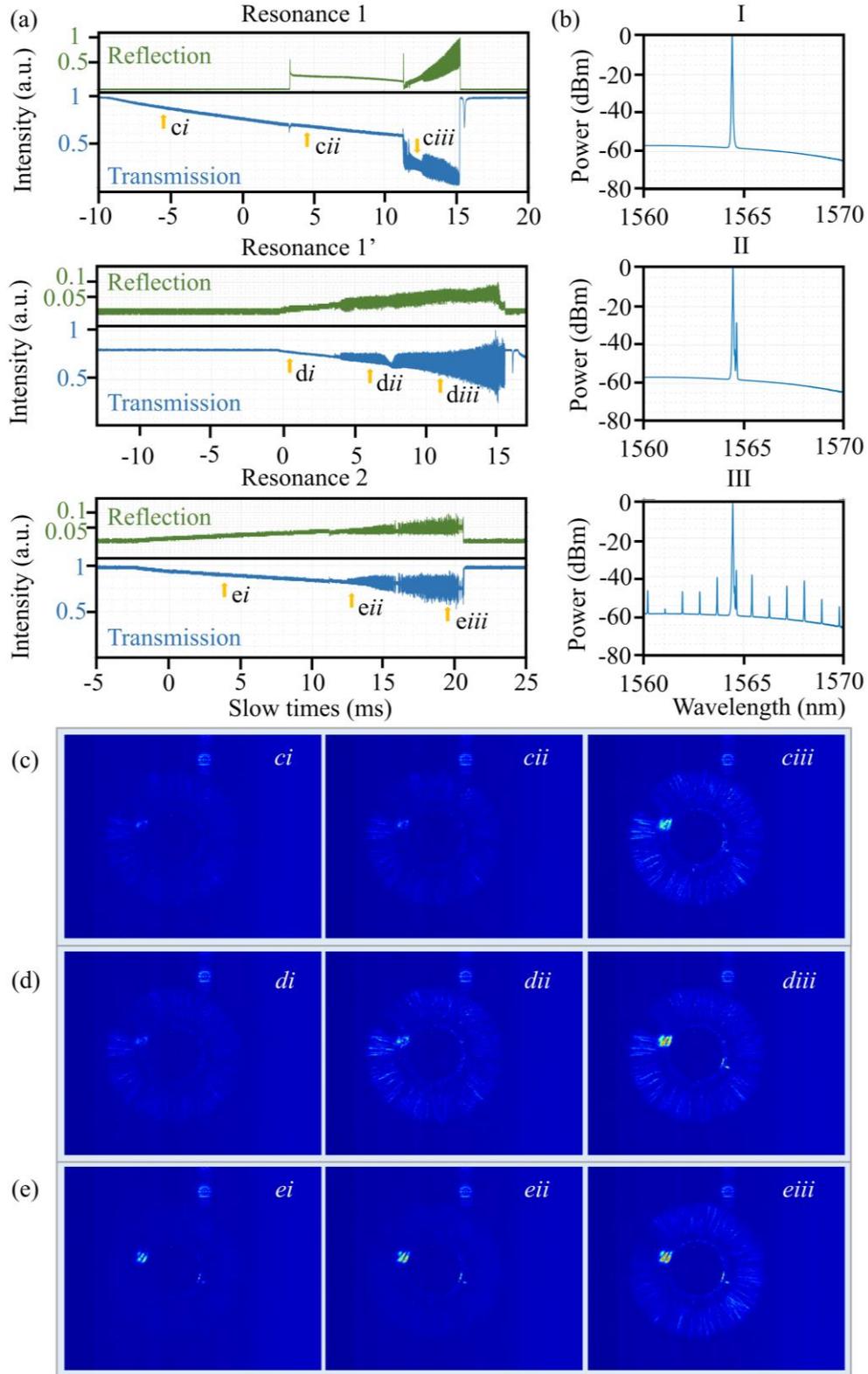

**Figure 4. Visualization of stimulated Brillouin scattering (SBS). (a)** Transmission (blue) and reflection (green) from two resonances that can generate SBS. **(b)** Optical spectra for Resonance 1 in three different states, labeled as i, ii, iii. Selected SWIR images for **(c)** Resonance 1, **(d)** Resonance 1', **(e)** Resonance 2. The sub-indices *i-iii* indicate different detuning values. Images recorded at the same transmitted powers are marked with the same sub-index.

To study the distribution and evolution of the scattered light, similar to the data in Fig. 2(j), we define an "inner" region (red ring) and an "outer" region (yellow ring). Then, we plot the scattering distribution along the resonator as a function of detuning in Fig. 5(a) for Resonance 1, in Fig. 5(b) for Resonance 1', and in Fig. 5(c) for Resonance 2. A spatial distribution analysis at a particular detuning is shown in the SI. Fig. 5(d) and Fig. 5(e) show the optical spectra corresponding to Fig. 5(a) at two different laser detuning positions which are marked with arrows and annotated with "d" and "e" in Fig. 5(a). Similar annotations are used for all the following panels. At the beginning of the detuning scan in Fig. 5(a,b,c), only the pump light is observed in all three cases. When further reducing the detuning in Resonance 1 and Resonance 1', SBS and FWM processes appear, and we can see an abrupt change in the scattered light. This is the detuning, at which nonlinear effects are starting (SBS+comb for Resonance 1 and SBS for Resonance 1'), marked as region I and II respectively. Because of different coupling positions of the tapered fiber, the Resonance 1' is experiencing higher losses. FWM is not observed in Resonance 1' and the corresponding scattering pattern, shown in Fig. 5(b), is more chaotic compared to Resonance 1 (Fig. 5a) in region II. The presence of these characteristic scattering patterns allows us to distinguish different nonlinear processes. In contrast to the scattering patterns of Resonance 1 and 1', the scattering patterns in Resonance 2 exhibit a different behavior, even though both Resonance 1 and Resonance 2 can generate SBS + comb within region II. We can still distinguish them using the spatial mode information from region II in Fig. 5(a) and 5(c). In addition, chaotic FWM (Fig. 5(h) in region II) and soliton generation (Fig. 5(i)) in region III can also be observed and distinguished.

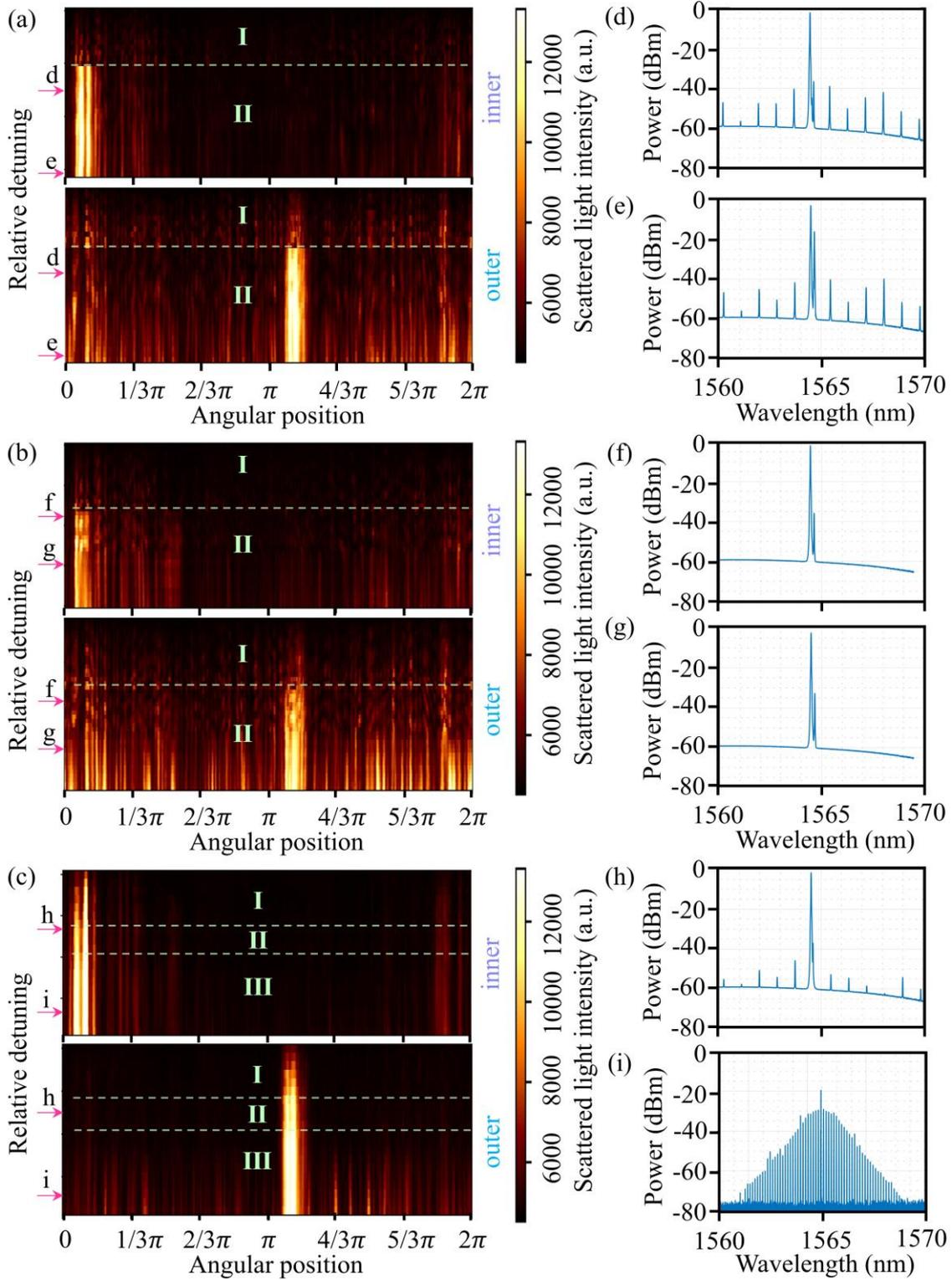

**Figure 5. Analysis of scattered light for various stimulated Brillouin scattering states.** Scattering intensity color maps as functions of angular position and detuning corresponding to **(a)** Resonance 1, **(b)** Resonance 1', **(c)** Resonance 2. The light green lines indicate position of the state changes observed on the optical spectrum analyzer. **(d)-(i)** Optical spectra measured at the detuning values highlighted with pink arrows on the color maps.

# Discussion and outlook

In this work, we show that scattered light can be used to characterize nonlinear dynamics within microresonators. This method can provide spatially resolved information on nonlinear processes in microresonators, in contrast to typically used indirect measurements through a bus-waveguide. In our experiments we can track the evolution of scattering patterns during different linear and nonlinear processes. The direct investigation of intra-cavity dynamics holds significant potential for both fundamental science and industrial applications. In fundamental science, it enables efficient identification and distinction of complex nonlinear phenomena. For industrial applications, the direct imaging of scattering patterns enables the development of real-time monitoring systems for integrated photonic circuits both during fabrication and during operation. In addition, the scattering patterns could be used to identify abnormalities in photonic structures and create physically unclonable optical chips [51]. Combining imaging techniques with machine learning-based classification systems can enable efficient identification of different nonlinear processes. Additionally, this method has potential applications in medical science, where it could visualize nonlinear effects induced by cells or complex molecules on microresonators.

# Author contributions

H.Z. and H.C.Z. performed the experiments. H.Z. fabricated the microresonator. H.Z., H.C.Z. and A.G. analyzed the data. A.G., S.Z., T.B., Y.Y.Z., L.H., J.X., A.P., Y. Y. Z. participated in the experiments. H.Z. and H.C.Z. wrote the manuscript, with input from all authors. P.D. supervised the project.

# Acknowledgement

This research was supported by the National Natural Science Foundation of China (62205145); National Key Research and Development Program of China (2022YFB2802700; Natural Science Foundation of Jiangsu Province (BK20220887); China Scholarship Council (202106830063); European Union's H2020 ERC Starting Grant "CounterLight" (756966); H2020 Marie Sklodowska-Curie COFUND "Multiply" (713694); Marie Curie Innovative Training Network "Microcombs" (812818); the University Development Fund (UDF01003527, The Chinese University of Hong Kong, Shenzhen); Shenzhen Science, Technology and Innovation Commission under Grant (JCYJ20240813113503005).